\documentclass[12pt,a4paper]{article}
\usepackage{amsmath}
\usepackage[dvips]{graphicx}
\input epsf
\usepackage{times}
\begin{document}
\begin{titlepage}
\title{Off-shell unitarity and geometrical
 effects in deep--inelastic scattering and  vector--meson electroproduction}
\author{S. M. Troshin and
 N. E. Tyurin\\[1ex]
\small  \it Institute for High Energy Physics,\\
\small  \it Protvino, Moscow Region, 142280, Russia}
\normalsize
\date{}
\maketitle

\begin{abstract}
Deep--inelastic scattering at low $x$ and elastic  vector meson
production are considered  on the basis
of the off-shell extension of the $s$--channel unitarity.
We discuss  behavior of the structure function
$F_2(x,Q^2)$ at low $x$ and the total cross--section of virtual
photon--proton scattering and obtain, in particular, the dependence
$\sigma^{tot}_{\gamma^* p}\sim (W^2)^{\lambda(Q^2)}$ where
exponent   $\lambda(Q^2)$  is related to the interaction radius of
a constituent quark. The energy dependence of the total cross--section of
$\gamma^*\gamma^*$--interactions is  calculated. The explicit mass
dependence of the exponent in the power energy behavior of the
vector meson production in the processes of virtual photon interactions
with a proton $\gamma^*p\to Vp$ has been obtained. We also consider angular
distributions at large momentum transfers.\\[2ex]
PACS Numbers: 12.40.Pp, 13.60.Hb, 13.60.Le
\end{abstract}
\end{titlepage}
\setcounter{page}{2}

\section*{Introduction}
Experimental data obtained at HERA  \cite{her}  clearly
demonstrated rising behavior of the structure function
$F_2(x,Q^2)$ at small $x$ which is translated to the rising
dependence of
 the  total cross--section
 $\sigma^{tot}_{\gamma^*p}(W^2, Q^2)$
 on center of mass energy $W^2$.
This effect
 is consistent with various $W^2$ -- dependencies
  and has been treated in different ways, e. g.  as a manifestation of
 hard BFKL Pomeron \cite{lipa}, a confirmation of the DGLAP evolution
 in  perturbative QCD \cite{pqcd},
 a transient phenomena, i.e. preasymptotic effects \cite{nad}
 or as a true asymptotical
dependence of the off--mass--shell scattering amplitude
\cite{petr}. This list is far from being complete and other
interpretations can be found, e.g. in the review papers \cite{her,lands}.

It is worth to  note here that
the  essential point in the study of low-$x$ dynamics
is that the space-time structure of
the scattering
at small values of $x$ involves  large distances
$l\sim 1/mx$ on the light--cone \cite{pas}, and the
region of $x\sim 0$ is  sensitive to the  nonperturbative contributions.
Deep--inelastic scattering in this region turns out to be a coherent
process where diffraction plays a major role and nonperturbative models
such as  Regge or vector dominance model can be competitive with perturbative
QCD and successfully applied for  description of the experimental data.

The
strong experimentally observed rise of
$\sigma^{tot}_{\gamma^*p}(W^2,Q^2)$ when
\begin{equation}\label{wlq}
\sigma^{tot}_{\gamma^*p}(W^2,Q^2)\propto (W^2)^{\lambda(Q^2)}
\end{equation}
with $\lambda(Q^2)$ rising with $Q^2$ from about $0.1$ to
$0.4$
 was considered to a somewhat extent as a surprising fact
on the grounds of our knowledge of
 energy dependence of the total cross--sections in hadronic
 interactions, where $\lambda\sim 0.1$.
The above comparison between photon--induced and hadron--induced
interactions is quite legitimate since the photon is demonstrating
its hadronlike nature for a long time.  The apparent difference
between the hadron and virtual photon total cross--sections
however has no fundamental meaning since there is no
Froissart--Martin bound in the case
off--shell particles \cite{petr,indur}. Only under some additional
assumptions this bound can be applied \cite{ttpre,levin}.

The most common form of unitarity solution -- the eikonal one --
 was generalized for
the off--shell scattering in \cite{petr}.
In this paper we consider
off--shell extension of the $U$--matrix approach to the amplitude
unitarization.  It is
shown that this approach along with
 the respective extension of the chiral quark model for the
 $U$--matrix \cite{csn}  leads to (\ref{wlq}), where
the exponent $\lambda (Q^2)$ is related to the $Q^2$--dependent
 interaction radius of an off--shell constituent quark.

It is to be stressed here the importance of the effective interaction
radius concept \cite{log}.  The study of the effective interaction radius
dependence on the scattering variables seemed very useful for
understanding of the relevant dynamics of high energy hadronic
reactions \cite{chy,khru}.
Now on it is widely known that the respective
geometrical considerations about interaction provide a deep insight
in hadron dynamics and deep--inelastic scattering

Besides  the studies of deep--inelastic scattering at low $x$
the interesting measurements of the characteristics
of the elastic
vector meson production were performed in the experiments H1 and
ZEUS at HERA \cite{zosa,melld}. It was shown that the integral  cross section
of the elastic vector meson production increases with energy in the way
similar to the
$\sigma^{tot}_{\gamma^*p}(W^2, Q^2)$ dependence
 on  $W^2$ \cite{her}.
It appeared that an increase of  the vector meson
electroproduction cross--section  with energy is  steeper for
heavy vector mesons as well as in the case when the virtuality
$Q^2$ is high. Discussion of such a behavior in  various model
approaches based on the nonperturbative hadron physics or
perturbative QCD can be found, e.g. in the  papers \cite{lands}.

It was already mentioned that we use an approach based on the
off-shell extension of the $s$--channel unitarity. Its application
to the elastic vector meson production in the processes
$\gamma^*p\to Vp$ allows in particular to consider angular
dependence and mass effects in these processes. It appears that
the obtained mass and $Q^2$ dependencies are in agreement with the
experimentally observed trends.  It is also valid for the angular
distribution at large momentum transfers.

\section{Off--shell unitarity}
The extension of the $U$--matrix unitarization for the off-shell
scattering was considered in \cite{ttpre}. It was supposed that
the virtual  photon fluctuates into a quark--antiquark
pair $q \bar q$ and this pair can
be treated as an effective virtual  vector meson state
in the processes with small Bjorken  $x$.
This effective virtual meson then interacts with a hadron. We
considered  a single effective vector meson field and used
for the amplitudes  of the processes
\begin{equation}
V^*+h  \to  V^*+h,\quad
V^*+h \to  V+h\quad \mbox{and}\quad
V+h \to  V+h
\end{equation}
the notations $F^{**}(s,t,Q^2)$, $F^{*}(s,t,Q^2)$ and $F(s,t)$ respectively,
i. e. we denoted that way the
amplitudes when both initial and final mesons  are off mass
shell, only initial meson is off mass shell and both mesons are on
mass shell.

The extended unitarity relation  for the amplitudes $F^{**}$ and
$F^*$ has a similar structure as the equation for the on--shell
amplitude $F$ but relate the different amplitudes. In  impact
parameter representation at high energies it  relates the
amplitudes $F^{**}$ and $F^*$ in the following way
\begin{equation}\label{offs}
\mbox{Im} F^{**}(s,b,Q^2) =  |F^{*}(s,b,Q^2)|^2+\eta^{**}(s,b,Q^2),
\end{equation}
where $\eta^{**}(s,b,Q^2)$ is the contribution to the unitarity of
many--particle intermediate on--shell states. The function
$\eta^{**}(s,b,Q^2)$ is the sum of the $n$--particle production
cross--section in the process of the virtual meson interaction
with a hadron $h$, i. e.
\[
\eta^{**}(s,b,Q^2)=\sum_n \sigma_n(s,b,Q^2).
\]
Similar relation exists for the functions $F^{*}$ and $F$.
It is worth noting that the solution of the off--shell unitarity
in the nonrelativistic case for a $K$--matrix representation
was obtained for the first time in \cite{lov}.
The solution of the off--shell unitarity can be written in the
$U$--matrix form \cite{ttpre}:
\begin{eqnarray}
F^{**}(s,b,Q^2) & = & U^{**}(s,b,Q^2)+iU^{*}(s,b,Q^2)F^{*}(s,b,Q^2)\nonumber\\
F^{*}(s,b,Q^2) & = & U^{*}(s,b,Q^2)+iU^{*}(s,b,Q^2)F^{}(s,b).\label{es}
\end{eqnarray}
The solution  of this system has a simple form when the following
factorization relation is supposed to be valid
\begin{equation}
[U^{*}(s,b,Q^2)]^2-U^{**}(s,b,Q^2)U(s,b)=0.\label{zr}
\end{equation}
Eq. (\ref{zr}) implies the following representation for the
functions $U^{**}$ and $U^{*}$:
\begin{eqnarray}
U^{**}(s,b,Q^2) & = & \omega ^2(s,b,Q^2)U(s,b)\nonumber\\
U^{*}(s,b,Q^2) & = & \omega(s,b,Q^2)U(s,b).\label{fct}
\end{eqnarray}
It is valid, e. g. in the Regge model with factorizable residues
and the $Q^2$--independent trajectory. It is also valid in the
off--shell extension of the chiral quark model for the $U$--matrix
which we will consider further. Thus, we  have for the amplitudes
$F^*$ and $F^{**}$
\begin{eqnarray}
F^{*}(s,b,Q^2) & = & \frac{U^{*}(s,b,Q^2)}{1-iU(s,b)}=
\omega(s,b,Q^2)\frac{U(s,b)}{1-iU(s,b)}
\label{vrq}\\
F^{**}(s,b,Q^2) & = & \frac{U^{**}(s,b,Q^2)}{1-iU(s,b)}=
\omega^2(s,b,Q^2)\frac{U(s,b)}{1-iU(s,b)} \label{vr}
\end{eqnarray}
and unitarity  provides inequalities
\begin{equation}
|F^*(s,b,Q^2)|  \leq  |\omega (s,b,Q^2)|,\quad
|F^{**}(s,b,Q^2)|  \leq  |\omega^2(s,b,Q^2)|.\label{bnd}
\end{equation}
It is worth noting that the above limitations are   much less stringent
than the limitation for the on--shell amplitude $|F(s,b)|\leq 1$.

 When  the function $\omega(s,b,Q^2)$ is real  we can write down
a simple expression for the inelastic overlap function
$\eta^*(s,b,Q^2)$:
\begin{equation}\label{etaf}
  \eta^{**}(s,b,Q^2)  =
\omega^2(s,b,Q^2)\frac{\mbox{Im} U(s,b)}{1-iU(s,b)}
\end{equation}

\section{Off--shell  scattering  in the $U$--matrix method}
In this section
 we consider  off--shell extension of the model for
hadron scattering \cite{csn}, which is based on the ideas of chiral quark
models.   Valence quarks located in the
central part of a hadron are supposed to scatter in a
quasi-independent way by the effective field.  In accordance with
that we represent the basic
dynamical  quantity in  the factorized form.
In the case   when the one of the hadrons
(vector meson in our case)  is off mass shell  the off--shell
$U$--matrix, i.e.  $U^{**}(s,b,Q^2)$ is represented as the
 product
\begin{equation} U^{**}(s,b,Q^2)\,=\, \prod^{n_{h}}_{i=1}\,
\langle f_{Q_i}(s,b) \rangle \prod^{n_{V}}_{j=1}\, \langle
f^{}_{Q^*_j}(s,b,Q^2) \rangle .\label{prdv} \end{equation}

 Factors $\langle
f_{Q}(s,b)\rangle$ and $\langle f_{Q^*}(s,b,Q^2)\rangle$
 correspond to the individual quark
scattering
 amplitude smeared over transverse position of the constituent quark
  inside hadron
and over fraction of longitudinal momentum of the initial
  hadron carried by this quark.
Under the virtual constituent quarks we mean the ones
 composing the virtual
 meson.
Factorization (\ref{prdv}) reflects the coherence in the valence
quark scattering and may be considered as an effective
implementation of constituent quarks' confinement.
The  picture of hadron structure with the valence constituent quarks
located in the central part and the surrounding condensate
implies  that     the
overlapping of hadron structures and interaction of the
condensates
 occur  at  the first stage of the collision.
 Due to an excitation of the condensates,
the quasiparticles, i.e. massive quarks arise. These quarks play role
of scatterers.
 To estimate number
of such quarks one could assume that  part of hadron energy carried by
the outer condensate clouds is being released in the overlap region
 to
generate massive quarks. Then their number can be estimated  by
the quantity:
 \begin{equation} \tilde{N}(s,b)\,\propto
\,\frac{(1-k_Q)\sqrt{s}}{m_Q}\;D^{h}_c\otimes D^{V}_c,
\label{4}
\end{equation} where $m_Q$ -- constituent quark mass, $k_Q$ -- hadron
 energy fraction
  carried  by  the constituent valence quarks. Function $D^h_c$
describes condensate distribution inside the hadron $h$, and $b$ is
an impact parameter of the colliding hadron $h$ and meson $V$.
Thus, $\tilde{N}(s,b)$ quarks appear in addition to $N=n_h+n_V$
valence quarks. Those quarks are transient
ones: they are transformed back into the condensates of the final
hadrons in elastic scattering. It should be noted that we use subscript
$Q$ to refer  the  constituent quark $Q$ and the same letter $Q$
is used to denote a virtuality $Q^2$. However, they enter formulas
in a way excluding  confusion.

The amplitudes $\langle f_Q(s,b)\rangle $ and
$\langle f_{Q^*}(s,b,Q^2)\rangle $ describe elastic
scattering  of a single
valence  on-shell  $Q$ or the off--shell  $Q^*$ quarks
with the effective field and we use for the function $\langle
f_Q(s,b)\rangle $  the following expression
\begin{equation}
\langle f_Q(s,b)\rangle =[\tilde{N}(s,b)+(N-1)] \,V_Q(\,b\,)
\label{7}
\end{equation}
where $V_Q(b)$  has a simple form
$V_Q(b)\propto g\exp(-m_Qb/\xi )$,
which corresponds to the quark interaction radius
$r_Q=\xi/m_Q$.
The function $\langle f_Q^*(s,b,Q^2)\rangle $ is to be written
as
\begin{equation}
\langle f^{}_{Q^*}(s,b,Q^2)\rangle =[\tilde{N}(s,b)+(N-1)] \,V_{Q^*}(b, Q^2).
\label{fqv}
\end{equation}
In the above relation
\begin{equation}
V_{Q^*}(b,Q^2)\propto g(Q^2)\exp(-m_Qb/\xi (Q^2) )
\end{equation}
and this form corresponds to the virtual constituent quark
interaction radius with effective field
\begin{equation}\label{rqvi}
r_{Q^*}=\xi(Q^2)/m_Q.
\end{equation}

The $b$--dependence of $\tilde{N}(s,b)$ is weak compared to the
$b$--dependence of $V_Q$ or $V_{Q^*}$ \cite{csn} and therefore we have taken
this function to be independent on the impact parameter $b$.
Dependence on virtuality $Q^2$ comes through dependence of the
intensity of the virtual constituent
 quark interaction $g(Q^2)$
and the parameter $\xi(Q^2)$, which determines the quark
interaction radius (in the on-shell limit $g(Q^2)\to g$ and
$\xi(Q^2)\to\xi$).
According to these considerations the explicit functional forms
for the generalized
reaction matrices $U^*$ and $U^{**}$
 can easily be written in the form of Eq. (\ref{fct}) with
\begin{equation}\label{omeg}
  \omega(s,b,Q^2)=\frac{\langle f^{}_{Q^*}(s,b,Q^2)\rangle}
  {\langle f^{}_{Q}(s,b)\rangle}.
\end{equation}
Note that Eqs. (\ref{zr}) and (\ref{fct}) imply that
\[
\langle f^{}_{Q^*\to Q}(s,b,Q^2)\rangle=
[\langle f^{}_{Q^*}(s,b,Q^2)\rangle \langle f^{}_{Q}(s,b)\rangle]^{1/2}.
\]

We consider the high--energy limit and for simplicity assume here
that all the constituent quarks have equal masses and parameters
$g$ and $\xi$ as well as $g(Q^2)$ and $\xi(Q^2)$
do not depend on quark flavor.
 We also asumme for simplicity
pure imaginary amplitudes. Then  the functions
$U$, $U^*$ and $U^{**}$ are the following
\begin{equation}
 U(s,b)  =  ig^N\left (\frac{s}{m^2_Q}\right )^{N/2}
\exp \left [-\frac{m_QNb}{\xi}\right ] \label{usb}
\end{equation}
\begin{equation}
U^*(s,b,Q^2)  =  \omega (b,Q^2) U(s,b),\quad
 U^{**}(s,b,Q^2)  =  \omega^2 (b,Q^2) U(s,b)\label{uvv}
\end{equation}
where the function $\omega$ is an energy-independent one
and has the following dependence on $b$ and $Q^2$
\begin{equation}\label{ome}
  \omega(b,Q^2)  =
  \frac{g(Q^2)}{g}\exp \left [-\frac{m_Qb}{\bar{\xi}(Q^2)}\right ]
\end{equation}
with
\begin{equation}\label{ksi}
  \bar{\xi}(Q^2)=\frac{\xi\xi(Q^2)}{\xi-\xi(Q^2)}.
\end{equation}

\section{Total cross--sections of $\gamma^* p$ and
 $\gamma^*\gamma^*$ interactions}
It is obvious that for
 the on--shell particles $\omega \to 1$ and
we  arrive  to the result obtained in \cite{ttpre} at large $W^2$
\begin{equation}\label{ons}
  \sigma^{tot}_{\gamma p}(W^2)\propto\frac{\xi^2}{m_Q^2}\ln ^2 \frac{W^2}{m_Q^2},
\end{equation}
where the usual for deep--inelastic scattering notation $W^2$
 instead of $s$ is used. Similar result is valid also for the off mass shell
 particles when the interaction radius of virtual quark does not depend
 on $Q^2$ and is equal to the interaction radius of the on--shell quark,
 i.e. $\xi(Q^2)\equiv \xi $. The behavior of the total cross--section
 at large $W^2$
\begin{equation}\label{ofs}
  \sigma^{tot}_{\gamma^* p}(W^2)\propto
  \left[\frac{g(Q^2)\xi}{gm_Q}\right]^2
  \ln ^2 \frac{W^2}{m_Q^2},
\end{equation}
 corresponds to the result obtained in \cite{ttpre}.
We consider further  the off-shell scattering with $\xi(Q^2)\neq\xi$
 and it should be noted
first that for the case when $\xi(Q^2)<\xi$
the total   cross--section would be energy-independent
\[
\sigma^{tot}_{\gamma^* p}(W^2)\propto\left[\frac{g(Q^2)\xi}
{g\lambda(Q^2)m_Q}\right]^2\]
 in the asymptotic region.
This
 scenario would  mean that the experimentally observed rise of
  $\sigma^{tot}_{\gamma^* p}$
 is transient preasymptotic phenomena \cite{nad,ttpre}.
  It can be realized when we
 replace in the formula for the interaction radius of the on--shell constituent
  quark $r_Q=\xi/m_Q$ the mass $m_Q$ by the value
   $m_{Q^*}=\sqrt{m_Q^2+Q^2}$ in order to obtain
  the interaction radius of the off-shell constituent quark and
  write it down as $r_{Q^*}=\xi/m_{Q^*}$, or equivalently
  replace  $\xi(Q^2)$ by
$\xi(Q^2)={\xi m_Q}/{\sqrt{m_Q^2+Q^2}}$.
The above
option cannot be excluded in principle.

However, when $\xi(Q^2)>\xi$ the situation is different
and we have at large
$W^2$
\begin{equation}\label{totv}
\sigma^{tot}_{\gamma^* p}(W^2,Q^2)\propto G(Q^2)\left(\frac{W^2}{m_Q^2}
\right)^{\lambda (Q^2)}
\ln \frac{W^2}{m_Q^2},
\end{equation}
where
\begin{equation}\label{lamb}
\lambda(Q^2)=\frac{\xi(Q^2)-\xi}{\xi(Q^2)}.
\end{equation}
We shall further concentrate  on this  we currently
think the most interesting case.

All the above expressions for $ \sigma^{tot}_{\gamma^* p}(W^2)$ can
be rewritten as the corresponding dependencies of $F_2(x,Q^2)$ at small $x$
according to the relation
\[
F_2(x,Q^2)=\frac{Q^2}{4\pi^2\alpha}
 \sigma^{tot}_{\gamma^* p}(W^2),
\]
where $x=Q^2/W^2$.

In particular,
(\ref{totv}) will appear in the  form
\begin{equation}\label{totv1}
F_2(x,Q^2)\propto\tilde{G}(Q^2)\left(\frac{1}{x}\right)^{\lambda (Q^2)}
\ln (1/x),
\end{equation}

It is interesting that the value and $Q^2$ dependence of the
 exponent $\lambda(Q^2)$ is related to the interaction radius
 of the virtual constituent quark. The value of parameter $\xi$
 in the model is determined by the slope of the differential cross--section
of elastic scattering at large $t$ \cite{lang}, i. e.
\begin{equation}\label{ore}
  \frac{d\sigma}{dt}\propto\exp\left[-\frac{2\pi\xi}{m_QN}\sqrt{-t}\right]
\end{equation}
and from the $pp$-experimental data it follows $\xi=2-2.5$. The uncertainty
is related to the ambiguity in the constituent quark mass value. Using for
simplicity $\xi=2$ and the data for $\lambda(Q^2)$ obtained
at HERA \cite{bart} we calculated the ``experimental'' $Q^2$--dependence
 of the function
$\xi(Q^2)$:
\begin{equation}\label{ksiq}
\xi(Q^2)=\frac{\xi}{1-\lambda(Q^2)}.
\end{equation}

The results are represented in Fig. 1. It is clear that experiment
 leads to $\xi(Q^2)$ rising with $Q^2$. This rise is slow and consistent with
 $\ln Q^2$ extrapolation. Thus, assuming this dependence to be kept
 at higher $Q^2$ and using (\ref{lamb}), we  predict saturation
 in the $Q^2$--dependence of $\lambda(Q^2)$, i.e. at large $Q^2$ the
 flattening will take place.
\begin{figure}[htb]
\vspace{2mm}
\begin{center}
\epsfxsize=4.1in \epsfysize=3.2in
 \epsffile{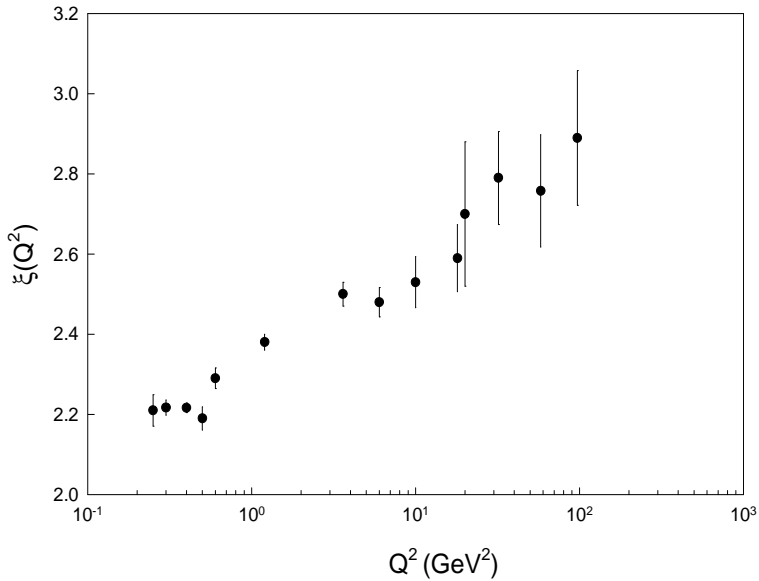}
\end{center}
 \caption[ksi]{ The ``experimental'' data for the function $\xi(Q^2)$.}
\label{fig:1}
\end{figure}

The above approach can be directly extended to the calculation of
the $\gamma^*\gamma^*$ total cross--section.
When  $\xi(Q^2)>\xi$ the following behavior of the
total cross--section
 at large
$W^2$ will take place:
\begin{equation}\label{totgg}
\sigma^{tot}_{\gamma^*\gamma^*}(W^2,Q_1^2,Q_2^2)
\propto G(Q_1^2)G(Q_2^2)\left(\frac{W^2}{m_Q^2}
\right)^{\lambda (Q_1^2)+\lambda (Q_2^2)}
\ln \frac{W^2}{m_Q^2}.
\end{equation}
This strong energy dependence of the $\gamma^*\gamma^*$--total cross--section
is consistent with  LEP data \cite{der}.
\section{Elastic vector meson production}

As it was already mentioned  we assumed that the virtual
photon before the interaction with the proton fluctuates into the
$\bar q q$ -- pair and for simplicity we limited ourselves with  light
quarks under discussion of the total cross--section.
The expression for the total cross-section is given by  (\ref{totv}).
The calculation of the the elastic and inelastic cross--sections can also be
directly performed in this approximation
using (\ref{usb}), (\ref{uvv}) and (\ref{ome}) and integrating
over  impact parameter
(\ref{vrq}) and (\ref{etaf}). Then we  obtain the following
dependencies for the cross--sections of elastic scattering and inelastic
interactions
\begin{equation}\label{elcrs}
\sigma^{el}_{\gamma^* p}(W^2,Q^2)\propto G_e(Q^2)\left(\frac{W^2}{m_Q^2}
\right)^{\lambda (Q^2)}
\ln \frac{W^2}{m_Q^2}
\end{equation}
and
\begin{equation}\label{inelcrs}
\sigma^{inel}_{\gamma^* p}(W^2,Q^2)\propto G_i(Q^2)\left(\frac{W^2}{m_Q^2}
\right)^{\lambda (Q^2)}
\ln \frac{W^2}{m_Q^2}
\end{equation}
with the universal exponent $\lambda (Q^2)$ given by the relation
(\ref{lamb}).
The above relations mean that the ratios of elastic and inelastic
cross--sections to the the total one are approximately constant
and do not depend on energy.

Now we  consider elastic (exclusive) cross--sections
both for  light and heavy
vector mesons production. We need to get rid of the light
quark limitation and extend the above approach in order to
include  the quarks with the different masses.
The inclusion, in particular, heavy vector meson production into this
scheme is straightforward: the virtual photon fluctuates before
the interaction with proton into the heavy quark--antiquark pair
 which constitutes
the virtual heavy vector meson state. After the interaction with a proton
this state turns out
into the real  heavy vector meson.

Integral exclusive (elastic) cross--section of vector meson production in
the process $\gamma^*p\to Vp$ when the vector meson in the final
state contains not necessarily  light quarks can be calculated directly
according to the above scheme and formulas of Section 2:
\begin{equation}\label{elvec}
\sigma^{V}_{\gamma^* p}(W^2,Q^2)\propto G_{V}(Q^2)\left(\frac{W^2}
{{m_Q}^2}
\right)^{\lambda_{V} (Q^2)}
\ln \frac{W^2}{{m_Q}^2},
\end{equation}
where
\begin{equation}\label{lavm}
\lambda_{V}(Q^2)= \lambda (Q^2)\frac{\tilde{m}_Q}{\langle m_Q \rangle}.
\end{equation}
In (\ref{lavm}) $\tilde{m}_Q$ denotes the mass of the constituent
quarks from
the vector meson and $\langle m_Q \rangle$ is the mean constituent
quark mass
of  system of the vector meson and proton system.
Evidently
$\lambda_{V}(Q^2)=\lambda(Q^2)$
for the light vector mesons.
This result is consistent with the most recent ZEUS
data, but statistics is still limited \cite{zosa}.
In the case when the vector meson
is very heavy, i.e. $\tilde m_Q\gg m_Q$ we
have
\[
\lambda_{V}(Q^2)=\frac{5}{2}\lambda(Q^2).
\]
We conclude that the respective cross--section
rises faster than the corresponding cross--section
of the light vector meson production, e.g. (\ref{lavm}) results in
\[
\lambda_{J/\Psi}(Q^2)\simeq 2\lambda(Q^2).
\]
This is in a qualitative agreement with the recently observed
trends in the HERA data \cite{melld}.

\section{Angular structure of elastic  and proton--dissociative
vector meson production}

Besides the integral cross--section of elastic vector meson
production it is interesting to consider angular distribution for these
processes. Recently the first measurements  of angular distributions
were performed \cite{critt} and it was found that angular
distribution in the light vector meson production is consistent
 with the power dependence $(-t)^{-3}$.

We apply the model described above for calculation of the differential
cross--sections in elastic vector meson production using an analysis
of the singularities of the amplitudes in the complex impact parameter plane
developed in \cite{lang}.

Since the integration goes over the variable $b^2$ rather than $b$
it is convenient to consider the complex plane of the variable
$\beta$ where $\beta=b^2$ and analyze singularities in the $\beta$--plane.
 Using  (\ref{vrq}) we can write down the
integral over the contour $C$ around a positive axis
in the $\beta$--plane:
\begin{equation}\label{cont}
  F^{*}(W^2,t,Q^2)= -i\frac{W^2}{2\pi^2}\int_C F^*(W^2,\beta,Q^2)K_0(\sqrt{t\beta})d\beta,
\end{equation}
where $K_0$ is the modified Bessel function and the variable $W^2$
was used instead of the variable $s$.
The contour $C$ can be closed at infinity
and the value of the integral will be then
determined by the singularities of the function
$F^*(W^2,\beta,Q^2)$, where
\[
F^*(W^2,\beta,Q^2)=\omega(\beta,Q^2)
  \frac{U(W^2,\beta)}{1-iU(W^2,\beta)}
\]
in a $\beta$--plane.

With explicit expressions for the functions $U$ and $\omega$ we conclude
that the positions of the poles in the complex $\beta$--plane are
\[
\beta_n(W^2)=\frac{\xi^2}{M^2}\{\ln \left[g^N\left (\frac{W^2}{{ m_Q}^2}
\right )^{N/2}\right]
+ i\pi n\}^2, \quad n=\pm 1,\pm 3,\ldots\; .
\]
where $M=\tilde{m}_Q n_V+m_Qn_h$. The location of the poles in the
complex impact parameter plane does not depend on the virtuality $Q^2$.
Besides the poles $F^*(W^2,\beta,Q^2)$ has a branching point at $\beta=0$
and
\[
\mbox{disc}\; F^*(W^2,\beta,Q^2)=
\]
\[
\frac{\mbox{disc} [\omega(\beta,Q^2)U(W^2,\beta)]-iU(W^2,\beta+i0)U(W^2,\beta-i0)
\mbox{disc}\; \omega(\beta ,Q^2)}{[1-iU(W^2,\beta + i0)][1-iU(W^2,\beta -i0)]},
\]
i.e.
\[
\mbox{disc}\; F^*(W^2,\beta,Q^2)\simeq i\;\mbox{disc}\; \omega(\beta,Q^2)
\]
since at  $W^2\to\infty$ the function $U(W^2,\beta)\to\infty$ at fixed $\beta$.
The function $F^{*}(W^2,t,Q^2)$  can be then represented as a sum of poles
and cut contributions, i.e.
\[
F^{*}(W^2,t,Q^2)=F_p^{*}(W^2,t,Q^2)+F_c^{*}(W^2,t,Q^2).
\]
The pole and
cut contributions are decoupled dynamically when $W^2\rightarrow
\infty  $. Contribution of
the poles determines the
amplitude  $F^{*}(W^2,t,Q^2)$ in the region $|t|/W^2 \ll 1$.  The
amplitude in this region can be represented in a form of  series:
\begin{equation}\label{polc}
F^*(W^2,t,Q^2)\simeq iW^2(W^2)^{\lambda_V(Q^2)/2}\sum_{n=\pm 1,\pm 3,\ldots}
\exp\left\{\frac{i\pi n}{N}\lambda_V(Q^2)\right\}\sqrt{\beta_n}
K_0(\sqrt{t\beta_n}).
\end{equation}
At moderate values of $-t$ when $-t \geq 1$ (GeV/c)$^2$ the amplitude (\ref{polc}) leads to
the Orear type behavior of the differential cross--section which is similar to
the Eq.(\ref{ore}) for the on--shell amplitude, i.e.
\begin{equation}\label{orev}
  \frac{d\sigma_V}{dt}\propto\exp\left[-\frac{2\pi\xi}{M}\sqrt{-t}\right].
\end{equation}

At small values of $-t$ the behavior of the differential cross--section
 is complicated,  the oscillating factors
$\exp\left\{\frac{i\pi n}{N}\lambda_V(Q^2)\right\}$ which
are absent in the on-shell scattering amplitude \cite{csn} play a role.

At large $-t$ the poles contributions is negligible  and contribution
from the cut at $\beta=0$ is a dominating one.
It appears that   the function $F_c^{*}(W^2,t,Q^2)$ does not depend  on energy
and differential cross section depends on $t$ in a power-like way
\begin{equation}\label{ttri}
  \frac{d\sigma_V}{dt}\simeq \tilde{G}(Q^2)\left(1-\frac{\bar{\xi}^2(Q^2)t}{\tilde{m}_Q^2}
  \right)^{-3},
\end{equation}
which is in  an agreement with the experimentally observed trends \cite{critt,melr}.
For large values of $-t$
\[
-t\gg \tilde{m}_Q^2/\bar{\xi}^2(Q^2)
\]
we have a simple $(-t)^{-3}$ dependence of the differential cross--section.
This dependence significantly differs from the one
in the on-shell
scattering \cite{csn} which approximates the quark counting rule \cite{matv}.
and this difference is in the large
extent because of the off-shell unitarity role.

The ratio of differential cross-sections for the production of the
different vector mesons
$\frac{d\sigma_{V_1}}{dt}/\frac{d\sigma_{V_2}}{dt}$  does
not depend on the variables $W^2$ and $t$ at large enough values of $-t$.

The production of the vector mesons accompanied by the proton dissociation into
the state $Y$ with mass $M_Y$
can be calculated  along the lines described in \cite{angd} with account for
 non-zero virtuality. The extension is straightforward.  Similar to the
case of on--shell particles we have a  suppression of the pole contribution at high
energies.
It is interesting to note that
the normalized differential cross-section
\[
\frac{1}{\sigma_0(W^2,M_Y^2,Q^2)}\frac{d\sigma }{ dtdM_Y^2}
\]
where $\sigma_0$ is the
value of cross-section at $t=0$ will exhibit a scaling behavior
\begin{equation}\label{scal}
\frac{1}{\sigma_0}\frac{d\sigma }{dtdM_Y^2}=
\left(1-\frac{4\xi^2t}{\tilde M^2(Q^2)}
\right)^{-3},
\end{equation}
where the $\tilde M$ is the following combination
\begin{equation}\label{ftau}
\tilde M(Q^2) =M_Y\left[1-\frac{2\tilde m_Q}{M_Y}\lambda(Q^2)\right].
\end{equation}
Note that $M(Q^2)\simeq M_Y$ at small values of $Q^2$ or when the value of $M_Y$
is large $M_Y\gg \tilde m_Q$.
The dependence (\ref{scal}) is in agreement with the experimentally observed dependencies
in the proton--dissociative vector meson production at large values of
 $t$ \cite{critt}.
\section{Conclusion}

We considered limitations the unitarity provides for the $\gamma^* p$--total
 cross-sections and geometrical effects in the
  model dependence of $\sigma^{tot}_{\gamma^* p}$.
In particular, it was shown that the constituent quark's
 interaction radius  dependence on $Q^2$  can lead to a nontrivial,
asymptotical result:
$\sigma^{tot}_{\gamma^* p}\sim (W^2)^{\lambda(Q^2)}$,
where $\lambda(Q^2)$ will be saturated at large values of $Q^2$.
This result is valid  when the interaction radius of the virtual
constituent quark is rising with virtuality $Q^2$. The reason
 for such rise
 might be of a dynamical nature and it could originate from
 the emission of the additional $q\bar q$--pairs in the
 nonperturbative  structure of a constituent quark.
In the present
approach constituent quark consists of a current quark
and the  cloud of quark--antiquark pairs of the different
flavors \cite{csn}.
Available experimental data
 are consistent with the $\ln Q^2$--dependence of the radius of this
 cloud .
The available experimental data for the structure functions
  at low values of $x$ continue to demonstrate the
  rising total cross-section of $\gamma^* p$--interactions
 and therefore we can consider it as a manifestation of
  the rising with virtuality
  interaction radius of a constituent quark. The steep energy increase
  of $\gamma^*\gamma^*$  total cross--section $ (W^2)^{2\lambda(Q^2)}$
   was also predicted.

We have also considered   the elastic vector meson production processes
in $\gamma^*p$--interactions. The mass and $Q^2$ dependencies of the integral
cross--section  of vector meson production
are related to the  dependence of the interaction radius  of the constituent
quark $Q$  on the respective quark mass $m_Q$ and virtuality
$Q^2$.
The behavior of the differential cross--sections at large $t$ is in the large
extent determined by the off-shell unitarity effects. These predictions
are in a qualitative agreement with  the experimental data.
The new experimental data
will be essential for the discrimination of the model approaches and studies of the
interplay between the non-perturbative and perturbative QCD regimes
(cf. e.g.  \cite{bart,cald}).

\section*{Acknowledgements} We are grateful to J. A. Crittenden  for
 the communications on the ZEUS experimental data on the angular distributions
 and to V. A. Petrov
 and A. V. Prokudin for  interesting discussions of the results.

\end{document}